\documentclass[%
twocolumn,
 amsmath,amssymb,
 aps,
 prl
]{revtex4-1}

\usepackage[utf8]{inputenc}

\usepackage{graphicx}
\usepackage{dcolumn}
\usepackage{bm}
\usepackage{graphicx}
\usepackage{amssymb,amsmath}
\usepackage{float}
\usepackage{longtable}
\usepackage{threeparttable}
\usepackage[toc,page]{appendix}
\usepackage[usenames, dvipsnames]{color}

\begin{document}

\title{Spin-Glass Charge Ordering in Ionic Liquids}

\author{Amir Levy[1], Michael McEldrew[2], Martin Z. Bazant[2,3]}
\affiliation{[1] Departments of Physics, Massachusetts Institute of Technology, Cambridge, Massachusetts 02139 USA}
\affiliation{[2] Departments of Chemical Engineering, Cambridge, Massachusetts 02139 USA}
\affiliation{[3] Departments of Chemical Engineering and Mathematics, Massachusetts Institute of Technology, Cambridge, Massachusetts 02139 USA}

\newcommand{\abf}{\mathbf{a}}
\newcommand{\pbf}{\mathbf{p}}
\newcommand{\kbf}{\mathbf{k}}
\newcommand{\qbf}{\mathbf{q}}
\newcommand{\xbf}{\mathbf{x}}
\newcommand{\rbf}{\mathbf{r}}
\newcommand{\ubf}{\mathbf{u}}
\newcommand{\dxx}{{\rm d}^3{\bf x}}
\newcommand{\drr}{{\rm d}^3{\bf r}}
\newcommand{\dpp}{{\rm d}^3{\bf p}}
\newcommand{\dqq}{{\rm d}^3{\bf q}}

\begin{abstract}
Ionic liquids form intricate microstructures, both in the bulk and near charged surfaces. In this Letter, we show that, given the ionic positions from molecular simulations, the ionic charges minimize a ``spin-glass" Hamiltonian for nearest-neighbor interactions with remarkable accuracy, for both room-temperature ionic liquids (RTIL) and water-in-salt electrolytes (WiSE). Long-range charge oscillations in ionic liquids thus result from positional ordering, which is maximized in ionic solids, but gradually disappears with added solvent. As the electrolyte becomes more disordered, geometrical frustration in the spin-glass ground state reduces correlation lengths. Eventually, thermal fluctuations excite the system from its ground state, and Poisson-Boltzmann behavior is recovered. More generally, spin-glass ordering arises in any liquid with anti-ferromagnetic correlations, such as 2D vortex patterns in super-fluids or bacterial turbulence.
\end{abstract}

\maketitle

{\it Introduction}- In recent years, room temperature ionic liquids (RTIL) have emerged as promising electrolytes for synthetic chemistry and electrochemical energy storage\cite{Fedorov2014,Kim2012,Scrosati2010,Garcia2004,Simon2008}. In the absence of solvent molecules, strong electrostatic interactions limit the applicability of classical mean-field approximations, such as the ubiquitous Poisson-Boltzmann (PB) theory \cite{Debye1923a} for dilute solutions. Extensions are available for steric effects \cite{Bazant2009a,Borukhov1997,Bikerman1942,Kilic2007,Kornyshev2007}, short ranged ion-ion forces \cite{Derjaguin1941,Verwey1948,Goodwin2017,Goodwin2017a,Adar2017,Gavish2018}, ion-solvent interactions \cite{Abrashkin2007,Gongadze2013} and Gaussian perturbations beyond mean-field \cite{Netz2000}, but no theory can fully describe the solvent-free limit of RTIL. 
\par
At electrified interfaces, ionic liquids share similarities with dilute electrolytes, and some aspects can be described by modified continuum models.  Direct surface force measurements reveal a diffuse electric double layer (EDL) structure, akin to that of a dilute aqueous solution\cite{Gebbie2013}, although the extent of this analogy is debated\cite{Lee2015}.  Nevertheless,  there have been some successful applications of mean-field continuum models to RTIL\cite{Kornyshev2007,Fedorov2010,Fedorov2014}, and strong electrostatic correlations, which induce charge ordering and oscillations\cite{Fedorov2008}, can be captured surprisingly well by higher-order PB type equations \cite{Santangelo2006,Hatlo2010,Bazant2011,Liu2013,Gavish2016a,Blossey2017}. 
\par
Strong charge correlations imply a non-local dielectric response, similar to that of polar solvents\cite{Kornyshev1978,Hildebrandt2004}. Bazant, Storey, and Kornyshev (BSK) extended the PB free energy functional to include both correlations and crowding effects and introduced the concept of a dielectric permittivity operator to approximate the non-local ionic polarization\cite{Bazant2011}. The BSK framework was subsequently used to describe a wide variety of structural \cite{Lee2013a,Yochelis2014,Moon2015,Liu2015b} and dynamical \cite{LeeAlphaA.andKondratSvyatoslavandVellaDominicandGoriely2015,Alijo2015,Jiang2016,Jiang2014,Stout2014} properties of ionic liquids and concentrated electrolytes. Yet, some phenomena, such as long-ranged under-screening\cite{Perez-Martinez2017,Smith2017,Perez-Martinez2017a} and charge-driven 3D structures of the double layer\cite{Ivanistsev2014,Ivanistsev2015,Rotenberg2015,Kornyshev2014}, are not captured by BSK or other continuum models, and coarse-grained charge profiles generally obscure correlated nano-structures\cite{Triolo2007,CanongiaLopes2006,Atkin2008}. 
\par
In this Letter, we use molecular dynamics (MD) simulations to reveal an essential and overlooked mechanism that determines the charge profile in ionic liquids:  geometrical frustration. Given the network of neighboring ionic positions in a symmetric binary mixture, we show that the charge distribution corresponds to the ground state of an effective spin-glass Hamiltonian \cite{Edwards1975}. We propose a minimization scheme based on a modified Goemans-Williamson (GW) algorithm\cite{Goemans1995} and perform spin-glass reconstructions of MD simulations of 1-Ethyl-3-methylimidazolium bis(trifluoromethylsulfonyl)imide (EMIM-TFSI), a commonly studied RTIL\cite{Palm2012,Largeot2008,Matsumoto2005,Seki2006}, and so-called ``water-in-salt" electrolytes (WiSE), recently introduced for Li-ion batteries~\cite{Suo2013,Suo2015a,McEldrew2018}.

{\it Theory. --}  The partition function of ionic liquids (neglecting non-idealities) can be written as a sum over all spatial configurations ($\left\{\rbf_i\right\}$) and valencies ($z_i$):
\begin{eqnarray}
    Z & = & \int \prod_{i=1}^{N} d\rbf_i \left[ \sum_{\{z_i\}} \exp \left(- l_B \sum_{i\neq j} \frac{z_i z_j }{|\rbf_i - \rbf_j|}\right)\right]
    \nonumber\\
    & = & \int \prod d\rbf_i Z_r[\left\{\rbf_i\right\}],
\end{eqnarray}
where $l_B= \beta e^2/\varepsilon$ is the Bjerrum length, $e$ the elementary charge, $\beta=(k_B T)^{-1}$ the inverse temperature, and $\varepsilon$ the dielectric constant of the medium. $Z_r$ is a reduced partition function that depends on the ionic positions. The full partition function is the thermodynamic average over all positional configurations of the reduced partition function. Alternatively, we use MD simulations to extract typical positional configurations. 
\par
The reduced partition function is similar to a spin-glass, with the following Hamiltonian:
\begin{eqnarray}
\label{ColoumbGlassHamiltonian}
    H  =  \frac{1}{2}\sum_{i \neq j} J_{i j} z_i z_j, \quad {\rm where} \quad J_{ij}  =  \frac{l_B}{|\rbf_i - \rbf_j|}.
\end{eqnarray}
In the dilute limit ($l_B\rightarrow 0$) the Debye-Huckel mean-field approximation becomes valid, at rather small salt concentrations ($<100$mM) for aqueous solutions ($l_B \approx 7$\AA). In the opposite limit ($l_B\rightarrow \infty$) relevant for ionic liquids, when the Bjerrum length is large compared to the ionic spacing, temperature induced charge fluctuations around the ground state are negligible, and the charge distribution is better approximated by minimizing the Coulomb energy. 
\par
Minimizing a spin-glass Hamiltonian is a well-known NP-complete problem \cite{Garey2002} that cannot be solved exactly. The difficulty lies in the ``Ising-like" constraint on the charges: $z_i=\pm 1$, which can be expressed efficiently via a matrix, $Z_{ij}=z_i z_j$. By construction, the rank of $Z$ equals $1$, and its diagonal is $Z_{ii} = 1$. The Hamiltonian, in terms of $Z$, is simply ${\rm Tr}(ZJ)$. Relaxing the constraint on the rank of $Z$, and letting it take a full rank, greatly simplifies the problem, and allows for a polynomial time solution. This is the celebrated Goemans-Williamson (GW) max-cut algorithm \cite{Goemans1995}. In our context, the GW algorithm can be interpreted as letting the spins rotate in an $N$-dimensional space, where $N$ is the total number of spins in the system \cite{Chandra2008}. 
\par
The GW algorithm steps are as follows: 1) Minimize ${\rm Tr}(ZJ)$ subject to $Z_{ii}=1$; 2) find the Cholesky decomposition of $Z$ ($Z=S S^{T}$); and 3) choose a plane in the $N$ dimensional space, and assign the $i^{\rm th}$ Ising spin a sign (charge) according to the side of the plane where the $N$ dimensional spin $S_{ik}$ lies. To solve the minimization problem, we use CVX, a package for specifying and solving convex programs \cite{Grant2008,Grant2013}.
\par
The GW algorithm can be applied to any pair-wise interaction, and interestingly, we find that fully connected systems yield poor results. Instead, a dramatic improvement is achieved by considering an effective Hamiltonian with only short-range interactions, such as the following (empirical) interaction between an ion and its $n^{\rm}$the nearest neighbor: 
\begin{eqnarray}
J_{n}^{\rm eff} = \begin{cases}
    {\rm e}^{-n} & \text{n=1\ldots5}\\
    0 & n>5.
  \end{cases}
\end{eqnarray}
Due to screening, ion-ion interactions are thus limited to only a handful of nearest neighbor pairs. 
We further update the results of the GW algorithm according to a "local electro-neutrality" condition, until convergence:
\begin{equation}
\label{signEqn}
    z_i = -{\rm sign} \left(\sum_{j\neq i} \frac{z_j l_B }{|\rbf_i -\rbf_j|}\right).
\end{equation}
Finally, the algorithm is accelerated by selecting the bisecting plane perpendicular to the first principal component of $S$.

{\it Results. --} Let us now apply the modified GW algorithm to test our main hypothesis, that the charge distribution is determined by the ground state of a spin-glass Hamiltonian, given the positional configuration. A useful starting point is to examine systems in complete disorder, by simulating hard-sphere liquids with different packing fractions (see details in the supplementary material). Fig.~2(b) shows the charge distribution around a central ion in the ground state. We notice an interesting trade-off between the distance of closest approach and over-screening. When ions are free to approach each other, it is almost always favorable for the nearest neighbor to be of opposite charge, regardless of other ions in the environment. Neighbors further away are much less correlated. As ionic radii increase, ions tend to be more evenly spaced and screening is shared by several neighbors; a longer ranged oscillatory structure emerges. 
\begin{figure}[h]
  \includegraphics[width=\columnwidth ]{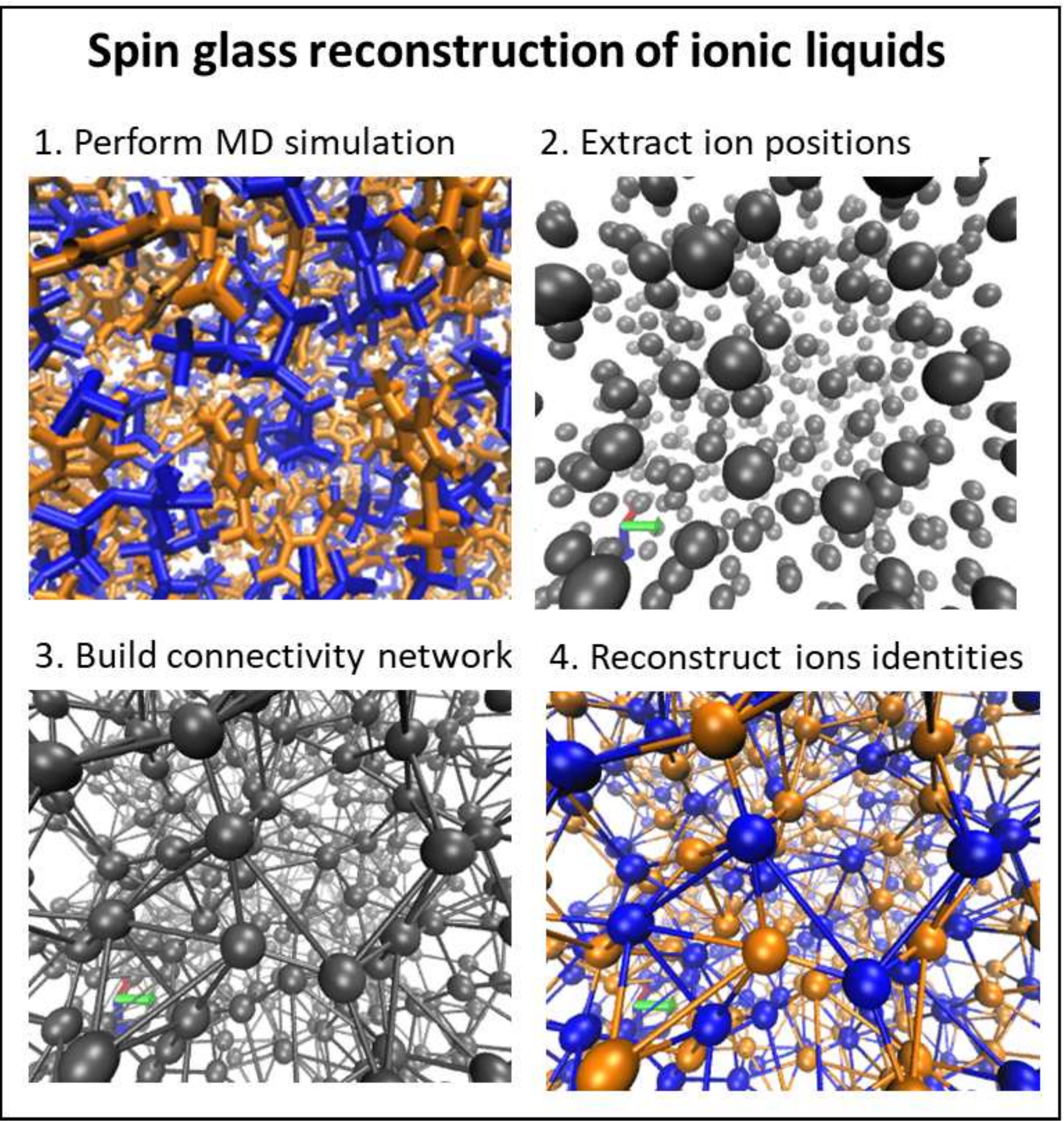}
  \caption{Illustration of the reconstruction procedure. 1 - the input is a full MD simulation of the ionic liquids. 2- The first step is to take a single snapshot, calculate the position of each molecule (as an average over its atomic positions), and delete the molecule identity. 3- Based on molecular positions, we construct a connectivity network, by connecting each molecule to its nearest neighbors. 4- Minimizing the spin-glass Hamiltonian for the network yields identities for the molecules, marked by orange and blue in the figure. The minimization is carried for each snapshot separately. }
  \label{fig1}
\end{figure}
\begin{figure*}[!]
  \includegraphics[width=\textwidth ]{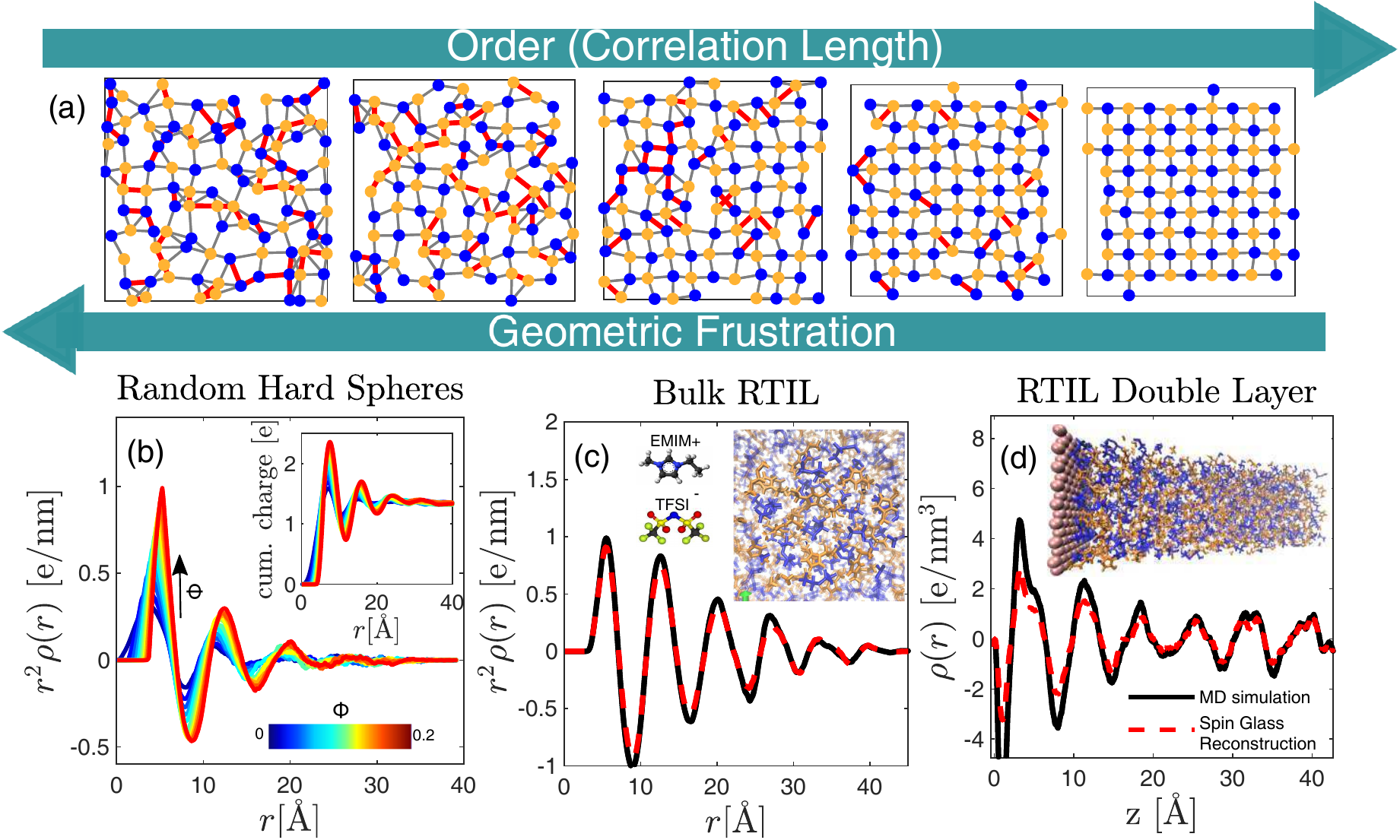}
  \caption{ Order vs. Frustration in Ionic Liquids. {\bf Top}- examples of 2D Ising model, with different degree of order. Red lines mark connections between parallel spins. The more ordered the system (right), more connections are satisfied. {\bf Bottom}- examples of 3D spin-glass with Coulomb interaction. (b)- Charge distribution around a central ion in a Random hard sphere model, for different packing fractions ($\Phi$), ranging from $0$ to $0.2$ in steps of 0.02.  (inset: cumulative charge distribution. Overscreening is defined as the maximum of this curve). (c) - Charge distribution around a central TFSI ion in EMIM-TFSI, based on MD simulation (black line) and the spin-glass reconstruction (dashed red) (inset: a snapshot from the MD simulation). (d) - EMIM-TFSI charge density near a weakly charged surface (0.01C/m$^2$), based on MD simulation (black line) and spin-glass reconstruction (dashed red) (inset: a snapshot from the MD simulation). } 
  \label{fig2}
\end{figure*}

Ionic liquids display a much longer correlation length. Data from scattering experiments, as well as MD simulations, reveals complicated nano-structures \cite{Triolo2007,Atkin2008,Gavish2018,Gavish2016} with oscillations that span many neighbors. We simulate an EMIM-TFSI ionic liquid to study these structures (see supplementary information for simulation details). As illustrated in Fig.~1, the Hamiltonian is constructed from ionic positions extracted from MD simulation snapshots. The minimization scheme is carried separately for each snapshot, and the results shown are averaged over all snapshots. Despite the complexity of the full atomistic MD simulations, the spin-glass model actually captures all the necessary physics: ionic valency almost exactly minimizes the Coulomb interactions. No other non-electrostatic ingredient is needed to recover the charge-density long ranged correlations.
\par
Fig.~2(c-d) compares results from MD simulations to the spin-glass reconstruction process. In bulk simulations, we recover the exact charge of almost $98\%$ of the ions. Consequently, the predicted charge distribution is almost indistinguishable from the simulated one (Fig.~2c). This exceptional match hints towards a unique ground state, and a high degree of order in the ionic positions. The reconstructed double-layer structure (Fig.~2d) fits reasonably well the simulated EDL, despite completely neglecting the interaction with the electrode. For weak surface charges, this interaction is only a secondary effect, but will have to be considered as electrode charge increases. A weakly charge electrode also exhibits a dramatic over-screening. The first layer can have a charge that is up to $15$ (!) times greater than the electrode charge. For comparison, BSK predicts an over-screening of only a few percent, which is more realistic for larger surface charges. 
\par
The spin-glass ground state aims to create long-ranged structures of alternating signs. Given the chance, a true long-range charge order would appear. Yet, this requires a high degree of order in the ionic positions. Even slight deviations from a perfect crystal structure lead to geometric frustrations: the pattern of alternating signs has to be broken in some direction (See Fig.~2(a) for illustration). In complete disorder, such pattern cannot exist at all, and correlations are limited to few neighbors only (Fig.~2(b)). In ionic-liquids, and especially near charged surfaces, a much more ordered structure appears, and facilitate large correlation lengths (Fig.~2(c-d)). 
\begin{figure}[h]
  \includegraphics[width=\columnwidth ]{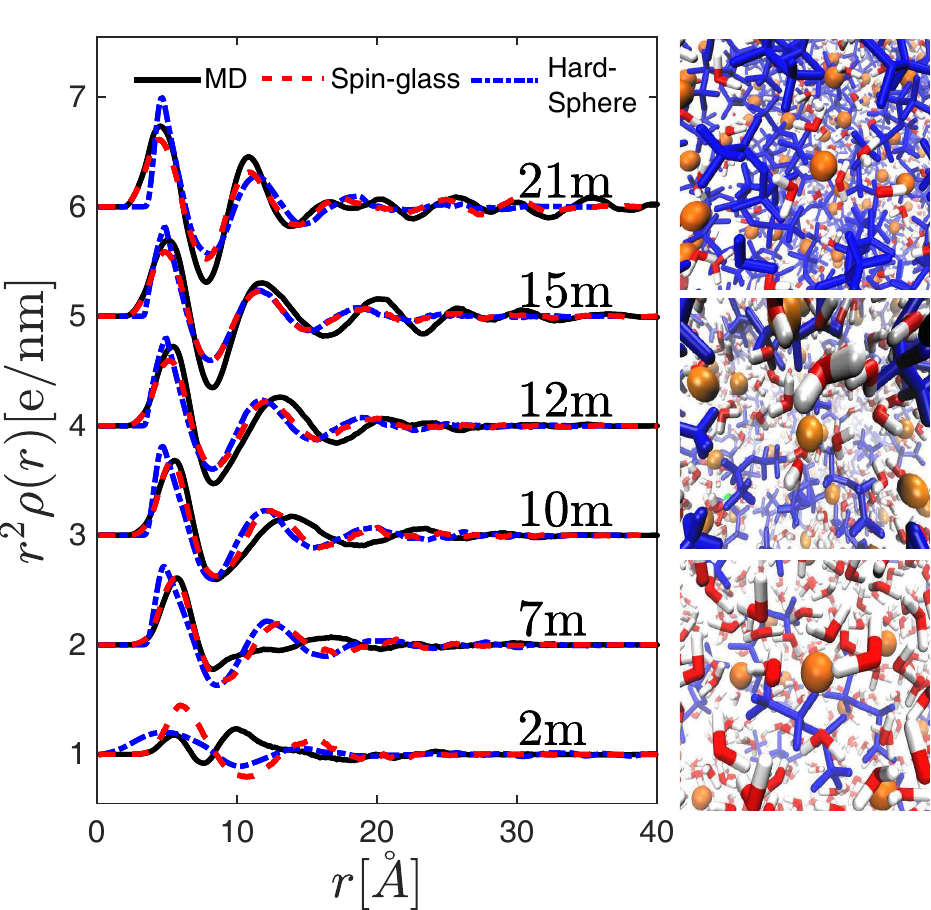}
  \caption{ LiTFSI in Water: from concentrated electrolyte to WiSE: Charge distribution around a central TFSI ion is shown for different molality ($1$ mol/kg solvent $= 1$m), from $2$m to $21$m. Each graph is plotted with an offset of $1$e/nm. Results from MD simulation (solid black line) are compared with random hard-sphere toy model (dashed-dot blue) and the spin-glass reconstruction (dashed red). Snapshots are shown from the LiTFSI MD simulation for $2$m(bottom), $7$m(middle) and $21$m (top). The hard sphere diameter equals $4.12$\AA.}
  \label{fig3}
\end{figure}
We conclude by examining MD simulations of a  Water-in-Salt Electrolyte (WiSE), with LiTFSI salt at varying concentrations. WiSE's are emerging as promising candidates to replace organic electrolytes in Lithium-ion Batteries\cite{Suo2013,Suo2015a,McEldrew2018}. They exhibit much shorter correlation lengths, even when the solvent concentration is small (Fig.~3). For moderate to high salt concentrations ($>5$mol/Kg), where ionic spacing is small compared to the Bjerrum length, our spin-glass framework is applicable. 
\par
Due to large size asymmetry, the spin-glass reconstruction only semi-quantitatively matches the simulations. The high molality limit ($21$m) is best reproduced by the minimization process, with about $80\%$ of ionic charges recovered. Similarly to RTIL, the hidden positional order stands behind this unique and easily accessible ground state. With increased water content ($7-15$m), the order gradually disappears, and we are only able to capture the general structure of the screening cloud. Upon decreasing ionic concentration further ($7$m and especially $2$m), thermal fluctuations triumphs and the spin-glass model breaks down. Yet, simple mean-field models are unsuitable for that regime as well, and ion-specific effects determine the correlation function. 
\par
When ionic positions are disordered, the charge distribution matches the random hard-sphere model (dashed blue lines in Fig.~3). Similarities are even more pronounced when only considering ordering relative to neighbour-number (Supplementary Fig.~S1). The reason for this high degree of disorder, compared with the RTIL, is twofold. First, there is a large positional entropy associated with small lithium ions, which is costly to suppress. Second, the solvent molecules weaken the electrostatic interactions. Maintaining a positional order is therefore unfavorable, and the WiSE resembles a hard-sphere liquid.
\par
{\it Discussion} The spin-glass model is a strong-coupling theory. It simplifies the complex interactions in ionic liquids and Water-in-Salt electrolytes to a minimization of a Hamiltonian with only local interactions (though corrections for electroneutrality are required). The correlation length is governed by geometric frustrations and increases with positional order. Such structures would emerge in any binary liquid with strong "anti-ferromagnetic" interactions and are not limited to Coulomb forces. Other examples include 2d vortex patterns that arise in super-fluids or bacterial turbulence\cite{Abo-Shaeer2001,Wioland2016} (see supplementary information for applying our scheme to bacterial vortexes). This is markedly different from the typical Debye-Huckel behavior, where electrostatic attraction competes with entropic "repulsion". 
\par
For solvent-free ionic liquids, the ground-state of the spin-glass Hamiltonian is easily accessible, and correlations are long-ranged. This might be the onset of a true long-range order in ionic-crystals. Room temperature ionic-crystals have much stronger interactions due to their small size, but we speculate that a similar regime of hidden positional order must exist, and play a role in the thermodynamics of melting. As solvent content increases, the energy landscape becomes more rugged, yet the system is still described well by its ground-state, and non-idealities are safely neglected. Eventually, in the moderately concentrated electrolyte regime ($<7$m), thermal fluctuations, as well as ion and solvent specific effects are dominating, and the spin-glass approach is no longer valid. 
\par
Though we do not offer here a general theory of ionic liquids and concentrated electrolytes, we believe our observations highlight the important physics. Any microscopic theory that wishes to describe the true nature of ionic liquids has to include the close interplay between charge and density ordering. Effective continuum models, on the other hand, might consider geometric frustration and positional ordering as some of the underlying microscopic driving forces. 
\par
This work was supported by a Professor Amar G. Bose Research Grant.

\bibliography{refs}

\pagebreak
\clearpage
\widetext
\begin{center}
\textbf{\large Supplementary Information}
\end{center}
\setcounter{equation}{0}
\setcounter{figure}{0}
\setcounter{table}{0}
\setcounter{page}{1}
\makeatletter
\renewcommand{\theequation}{S\arabic{equation}}
\renewcommand{\thefigure}{S\arabic{figure}}

\section{Simulations Details}
\subsection{MD simulations}
In this study, we performed all-atom classical MD simulations using LAMMPS \cite{plimpton1995}. We performed a set of simulations for two different systems: one set for a neat ionic liquid (IL), EMIM-TFSI and another for the Water-in-Salt Electrolyte (WiSE), LiTFSI (at varying concentrations). EMIM-TFSI was simulated in both full periodic geometries in order to study bulk-like properties, as well as nano-slit (slab) geometries in order to study the electrical double layer.  For the nano-slit geometry, we prescribe a constant surface charge density of $\pm0.1 \,C/m^2$ at two electrodes that sandwich the ionic liquid. LiTFSI electrolyte was simulated in fully-periodic geometries each set of simulations at molal concentrations of 2m, 7m, 10m, 12m, 15m, and 21m.  
\par
\emph{Simulation Details:} For EMIM-TFSI in the periodic geometries, we performed simulations containing 300 ion pairs. For the aqueous LiTFSI systems we performed simulations containing 1000 water molecules and enough ion pairs to make 2m, 7m, 10m, 12m, 15m, and 21m solutions. The simulations were performed at fixed temperature (300 K) and pressure (1 bar), with Nose-Hoover thermostat and barostat until the density of the fluid relaxed to a constant, which required 12 ns, with 1 fs time steps. Next, we switched to constant volume simulation box still with a fixed temperature of 300 K, again using the Nose-Hoover thermostat, and equilibrate for an additional 6 ns. Finally, production runs were performed for an additional 6 ns. The initial configurations for all simulations were generated using the open-source software, PACKMOL \cite{martinez2009packmol}. MD simulations were visualized using the open-source software, VMD \cite{HUMP96}.
\par
In the nano-slit geometry, we simulated the system at constant volume and temperature, filling a 33x33x200 ${\rm \AA}^3$ simulation box, with two 33x33x33 ${\rm \AA}^3$ gold electrodes (fcc lattice) sandwiching the electrolyte fluid, which fills the box at densities determined from the periodic simulations. The box contained 528 ion pairs and 4096 gold atoms. Surface charges were applied by placing partial charges on the first atomic layer (128 atoms) of gold, according to the specified surface charge density of $\pm 0.01 \,C/m^2$.  Equilibration runs of about 12 ns were performed initially with no applied potential/charge, with 1 fs time steps. Then the surface charge was stepped up from zero, allowing for 12 ns of equilibration and 6 ns of production at the $\pm 0.01 \,C/m^2$ electrode surface charge.
\par
\emph{Force Field Details}: For all ionic species, we employed the CL$\&$P force field, which was developed for ionic liquids, with same functional form as the OPLSAA force field\cite{CanongiaLopes2012}. For water, we employed the spc/e force field\cite{berendsen1987}. Interatomic interactions are determined using Lorentz-Berthelot mixing rules. Finally, for nano-slit simulations, we require force fields for the gold electrode. We did not explicitly model the dynamics of the electrode, omitting the need for a gold-gold force field. The gold was made to interact with the fluid mainly via Coulomb interactions, as the surface layer of gold atoms are charged according to the prescribed surface charge density. We also include Lennard-Jones interactions, which were made to be the same no matter what atom is interacting with gold (LJ well depth: $\varepsilon = 0.001\text{eV}$, LJ radius: $\sigma = 3 \AA$). We made the Lennard-Jones parameters constant for all species so that conclusions from the simulations that are not specific to the choice of the electrode material. Long range electrostatic interactions were computed using the Particle-Particle Particle-Mesh (PPPM) solver (with a cut-off length of 12 $\AA$), which maps particle charge to a 3D mesh for the periodic simulations and a 2D mesh in the transverse direction for the nano-slit simulation\cite{hockney1988}.

\begin{figure*}[h]
  \includegraphics[width=\textwidth ]{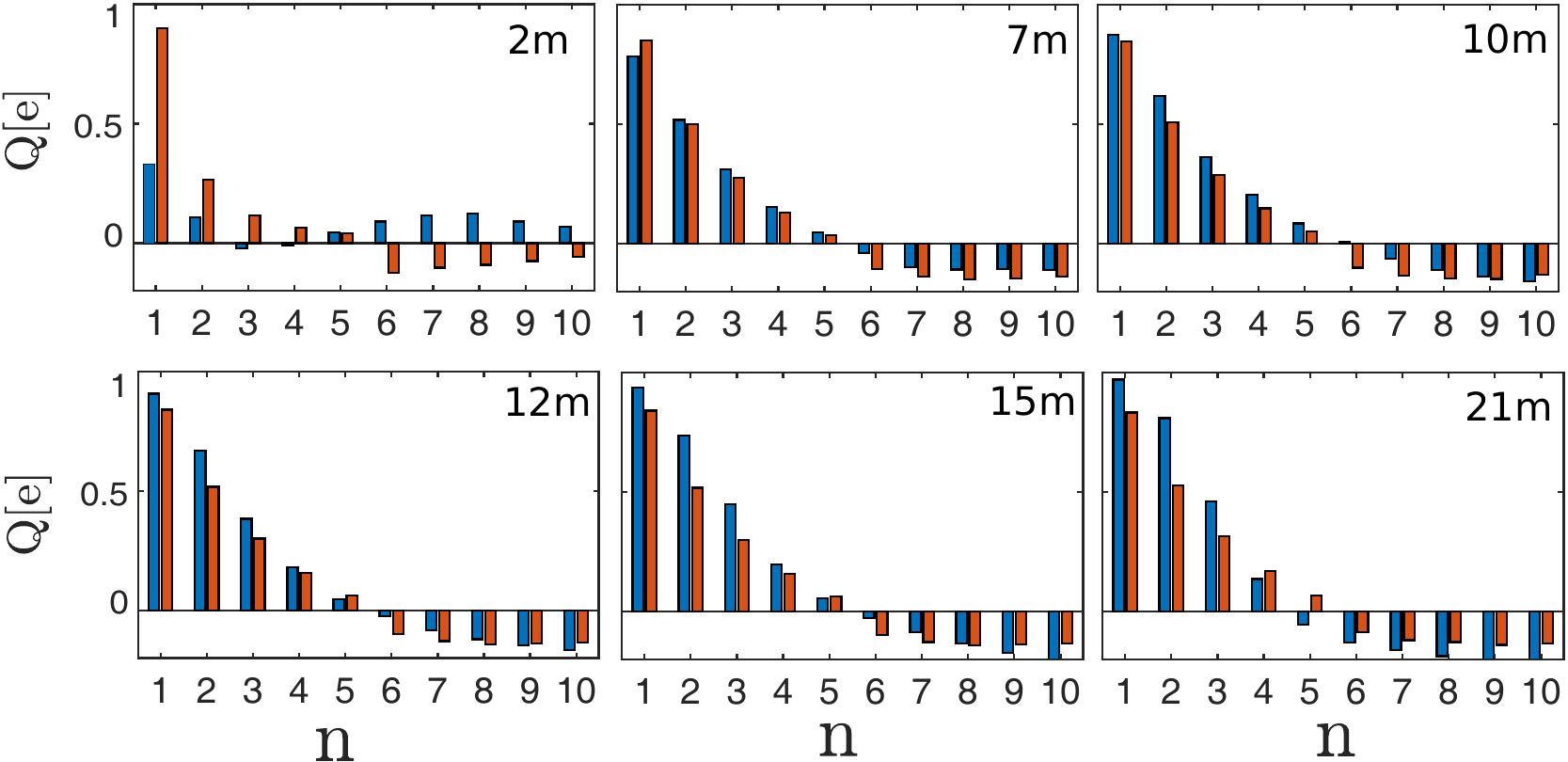}
  \caption{Screening cloud around a central TFSI ion is shown for different molality ($1$ mol/kg solvent $= 1$m), from $2$m to $21$m. Results from MD simulation (right blue bars) are compared with random hard-sphere toy model (left red bars). The hard sphere diameter equals $4.12$\AA.}
  \label{figA1}
\end{figure*}
\subsection{MC Simulations of random hard-sphere liquids}

In the main text, we describe a completely disordered spin-glass, based on a hard-sphere Monte-Carlo (MC) fluid simulations. For each packing fraction ($\Phi=0\ldots0.2$ in intervals of $0.02$), we simulated a system with $400$ particles. The size of the system was $50 \times 50 \times 50 {\rm \AA}^3$ with periodic boundary conditions, and the particles radii were set according to the desired packing fraction. Each simulation included $4000$ Monte-Carlo sweeps (moves/particle). Particles displacements were generated by moving the particle in a random orientation, with a normally distributed radius with standard deviations of $0.5$\AA. The move was accepted if it did not result in overlapping spheres. A snapshot was saved for $10\%$ of the sweeps. We used a random sequential addition (RSA) algorithm for the initial configuration. In this algorithm, particles are randomly placed in a periodic box. In each iteration, a new particle is considered and added if it does not overlap with any other particle already in the box. 

\section{Water in Salt Electrolytes Screening Cloud}

 The spin-glass reconstruction scheme captures the general charge ordering structure, but not the fine details. The precise location of the nearest neighbor, for example, is determined by van der Waals interactions rather than electrostatics. Hence, the charge distribution around a central ion is expected to depend on the simulated system. Conversely, the average charge as a function of the neighbor number, which depends on the topology of the network, has a more universal behavior. Fig.~S1 shows the first $10$ neighbors for both random-hard sphere model and aqueous LiTFSI simulation, for similar water content as Fig.~3 in the main text. A reasonable match is observed, especially in moderate concentrations ($7-15$m) where disorder is strongest.

\section{Turbulence in bacterial suspensions}

To demonstrate the generality of the spin-glass reconstruction scheme, we consider turbulence in bacterial suspensions as an example of a different disordered system with strong anti-ferromagnetic interactions. A Bacterial colony of Bacillus subtilis self-organizes into collective movement, and forms vortices under confinement\cite{dunkel2013}. To minimize drag forces and reduce friction adjacent vortices prefer to rotate in opposite directions. The details of this interaction follow complicated hydrodynamic equations, but as long as the anti-ferromagnetic interaction are strong, "spin" ordering is expected to dominate the emerging structure. We study the system with an effective spin-glass Hamiltonian, where the vortex directionality plays the role of spin, and the positions of the vortices cores are extracted from simulations. We use simulation data of swimming bacteria, adapted from \cite{dunkel2013}. $23$ core positional were extracted manually from a snapshot image the simulated flow field (Fig.S2-a). 

We arbitrarily choose the same form of local interactions as the effective RTIL Hamiltonian (Eq.~4 in the main text) but restrict connectivity only to vortices that are in physical contact via Delaunay triangulation. The minimization process was carried out using the modified GW algorithm, omitting the last stage of requiring electro-neutrality. Out of the $23$ vortices, the directionality of $19$ of them was reconstructed correctly (Fig.~S2-b). To illustrate the reconstructed vorticity (Fig.~S2-c), we superimpose a Lamb-Oseen (Gaussian) vortex at each core location, with angular velocity $\Omega(r)\propto \left[1-\exp[- (r/r_m)^2]\right]/r^2$, and a radius of $r_m=25\mu$m. The nice qualitative match illustrates the universality of our approach. The emerging structures in disordered anti-ferromagnetic systems are governed by the geometry, and are insensitive to details of their physical origin.

\begin{figure*}[h]
  \includegraphics[width=\textwidth ]{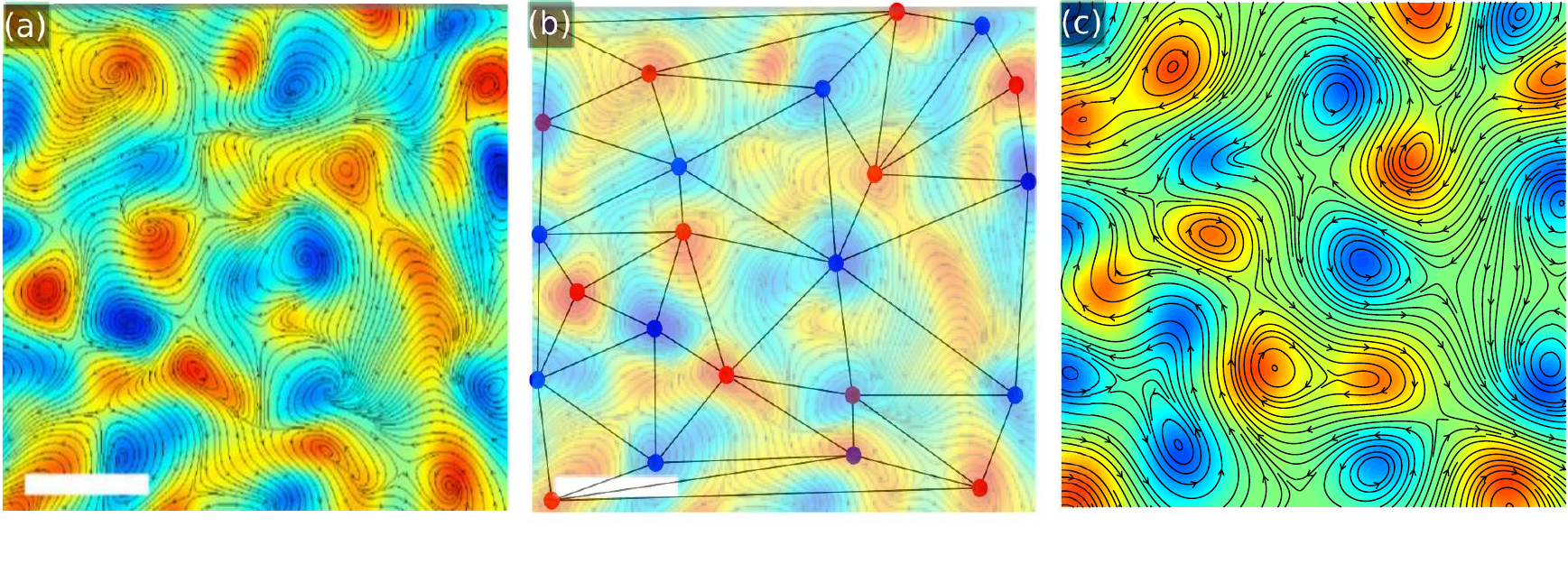}
  \caption{Applying spin-glass reconstruction for swimming bacteria. (a)- snapshot of simulated flow field (adapted from \cite{dunkel2013}, with permission) . (b)- A vortex network constructed by extracting the vortices centers as nodes. The sign of each node, clockwise (red) or counter-clockwise (blue) rotation, was derived from minimizing spin-glass Hamiltonian and matches $19$ out of the $23$ nodes of the simulation. (c)- Illustration of the reconstructed flow field, based on a superposition of independent Lamb–Oseen vortices}
  \label{figA2}
\end{figure*}

\end{document}